\documentclass[12pt] {article}
\usepackage{latexsym}

\newcommand{\beq}{\begin{equation}}
\newcommand{\feq}[1]{\label{#1} \end{equation}}
\newcommand{\beqr}{\begin{eqnarray}}
\newcommand{\feqr}{\end{eqnarray}}
\def\non{\nonumber}
\def\noi{\noindent}
\newcommand{\rf}[1]{~(\ref{#1})}

\def\ap#1#2#3{Ann. of Phys. {\bf{#1}} (#2) #3}

\def\grg#1#2#3{J. Gen. Rel. Grav. {\bf{#1}} (#2) #3}

\begin{document}

\renewcommand{\thefootnote}{\fnsymbol{footnote}}

\begin{center}

\vspace{2cm}

{\LARGE \bf A note on Riemann normal coordinates.}\\[6mm]

\large{Agapitos Hatzinikitas}\footnote [2] 
{e-mail: ahatzini@cc.uoa.gr}\\[8mm]

{\it University of Athens, \\
Nuclear and Particle Physics Division,\\
Panepistimioupoli GR-15771 Athens, Greece.}\\[8mm]

{\small \bf Abstract}\\[3mm]

\parbox{6.2in}
{\small The goal of this note is to provide a recursive algorithm that allows one to calculate the expansion of the 
metric tensor up to the desired order in Riemann normal coordinates. We test our expressions up to fourth order and predict 
results up to sixth order. For an arbitrary number of symmetric partial derivatives acting on the components of the 
metric tensor subtle treatment is required since the degree of complication increases rapidly.}

\end{center}


\section{Introduction}

Riemann normal coordinates (abbreviated as RNC from now on) have a vast field of applications to Physics. In particular 
they constitute the main ingredient of the background field method \cite{alvarez} widely used for nonlinear $\sigma$-
model calculations in curved spacetime. They have the appealing 
feature that the geodesics passing through the origin have the same form as the equations of straight lines passing 
through the origin of a Cartesian system of coordinates in Euclidean geometry. 

The purpose of this letter is to bridge the gap one confronts when attempts to perform computations that involve expansion 
of the metric tensor beyond the fourth order \cite{petrov}. We present the general method for breaking the barrier 
of the fourth order and at the same time we supply the reader with formulae adequate to reach the order of its choice.  

The starting point will be a geodesic, on a compact n-dimensional Riemannian manifold, parametrized by $x^{l}(s)$ 
and satisfying the differential equation:
\beqr
\frac{d^2 x^{l}}{ds^2} + \Gamma^{l}_{jk}(x) \frac{d x^{j}}{ds} \frac{d x^{k}}{ds} =0
\label{geoeq}
\feqr

\noi where s is the arc length and $\Gamma^{l}_{jk}$ denotes the Christoffel symbol for the Levi-Civita connection. 
Any integral curve of\rf{geoeq} is 
completely determined by a point $\mathcal{P}(x^{1}_{0},x^{2}_{0}, \cdots , x^{n}_{0})$ and a direction defined by 
the tangent vector $\xi^{l}= \left(\frac{dx^l}{ds}\right)_{\mathcal{P}}$. The power series solution of\rf{geoeq} is:
\beqr
x^{l}(s) = x^{l}_{0}+\sum_{k=1}^{\infty} \frac{1}{k!} \left( \frac{d^k x^l}{ds^k} \right)_{\mathcal{P}} s^k
\label{powe}
\feqr

\noi and its coefficients can be replaced using successive differentiations by:
\beqr
\left(\frac{dx^l}{ds} \right)_{\mathcal{P}} &=& \xi^l \non \\
\left(\frac{d^2 x^l}{ds^2} \right)_{\mathcal{P}} &=& -\left(\Gamma^{l}_{i_1 i_2} \right)_{P} \xi^{i_1} \xi^{i_2} \non \\
\left(\frac{d^3 x^l}{ds^3} \right)_{\mathcal{P}} &=& - \left[\partial_{(i_1} \Gamma^{l}_{i_2 i_3 )} 
- 2 \Gamma^{l}_{(i_1 \rho} \Gamma^{\rho}_{i_2 i_3 )} \right]_{P}  \xi^{i_1} \xi^{i_2} \xi^{i_3}  \non \\
\left(\frac{d^4 x^l}{ds^4} \right)_{P} &=& - \Big[ \partial_{(i_1} \partial_{i_2} \Gamma^{l}_{i_3 i_4 )}
- \partial_{\rho} \Gamma^{l}_{(i_1 i_2} \Gamma^{\rho}_{i_3 i_4 )} 
- 4\partial_{(i_1} \Gamma^{l}_{i_2 \rho} \Gamma^{\rho}_{i_3 i_4 )}
- 2 \partial_{(i_1} \Gamma^{\rho}_{i_2 i_3} \Gamma^{l}_{\rho i_4 )} \non \\
&+& 4 \Gamma^{l}_{k (i_1}\Gamma^{k}_{i_2 \rho} \Gamma^{\rho}_{i_3 i_4 )}
+ 2 \Gamma^{l}_{\rho \sigma} \Gamma^{\rho}_{(i_1 i_2}\Gamma^{\sigma}_{i_3 i_4 )} \Big]_{P}  
\xi^{i_1} \xi^{i_2} \xi^{i_3} \xi^{i_4}. \non \\
\cdots
\label{coeg}
\feqr

\noi where dots in\rf{coeg} represent higher derivative terms at the point $\mathcal{P}$ and symmetrization exclusively 
relates the lower i indices \footnote{Proof of the fourth expression in\rf{coeg} would require the use of the identity 
$\frac{dx^{\mu}}{ds} D_{\mu} \frac{dx^{\nu}}{ds} =0$}. Apparently the domain of 
convergence of\rf{powe} depends on the components  $g_{ij}$ and the values of $\xi^i$. However, for sufficiently 
small values of s it defines an integral curve of\rf{geoeq}.

\section{The computational procedure}

In RNC the geodesics through $\mathcal{P}$ are straight lines defined by:
\beqr
y^l=\xi^l s.
\label{strai}
\feqr

\noi This set of coordinates cannot be used to cover the whole manifold and is valid only for a small neighbourhood of 
the point $\mathcal{P}$ where the conditions of the existence and uniqueness theorem of differential 
equation of the geodesics are satisfied. In this region no two geodesics through $\mathcal{P}$ intersect due to 
one-one correspondence of $x^l$ and $y^l$. Incidently, this displays the local nature of RNC. 

Substituting\rf{strai} into eq.\rf{powe} one has:
\beqr
x^l = x^{l}_{0}+\sum_{k=1}^{\infty} \frac{1}{k!} \left(\Gamma^{l}_{i_1 i_2 \cdots
i_k}\right)_{\mathcal{P}} y^{i_1} y^{i_2} \cdots y^{i_k}
\label{subst}
\feqr

\noi where $\left(\Gamma^{l}_{i_1 i_2 \cdots i_k}\right)_{\mathcal{P}}$ are the 
$\textit{``generalized Christoffel symbols''}$ 
at the point $\mathcal{P}$ used in eq.\rf{coeg}. The Jacobian $\left|\frac{\partial x^l}{\partial y^m} \right|_{\mathcal{P}} 
\neq 0$ and thus the series\rf{subst} can be inverted.

For this local system of coordinates the geodesic equation can be written as:
\beqr
\bar{\Gamma}^{l}_{ij}(y) y^i y^j =0
\label{newgeo} 
\feqr

\noi and the power series solution becomes:
\beqr
y^l = \sum_{k=1}^{\infty} \frac{1}{k!} 
\left(\bar{\Gamma}^{l}_{i_1 i_2 \cdots i_k} \right)_{\mathcal{P}} 
\xi^{i_1} \xi^{i_2} \cdots \xi^{i_k} s^k.
\label{riepo}
\feqr

\noi Reduction of\rf{riepo} to\rf{strai} for arbitrary values of $\xi^l$ implies that at the origin $\mathcal{P}$ holds:
\beqr
\bar{\Gamma}^{l}_{(i_1 i_2 \cdots i_k )} =0
\label{syml}
\feqr

\noi or, by induction, one can easily prove that eq.\rf{syml} is equivalent to:
\beqr
\partial_{(i_1} \partial_{i_2} \cdots \partial_{i_{k-2}} 
\bar{\Gamma}^{l}_{i_{k-1} i_{k} )} =0
\label{eqre}
\feqr

\noi Paraphrasing eq.\rf{eqre} means that all symmetric derivatives of the affine connection vanish at the origin in RNC.
 
Generally speaking, a covariant second rank tensor field on a manifold can be expanded according to:
\beqr
T_{k_1 k_2}(\tilde{\phi})=\sum_{n=0}^{\infty} \frac{1}{n!} \left[
\frac{\partial}{\partial \xi_{i_1} } \frac{\partial}{\partial \xi_{i_2}}
\cdots \frac{\partial}{\partial \xi_{i_n}}\right] T_{k_1 k_2}(\phi)
\xi_{i_1}\xi_{i_2} \cdots \xi_{i_n}.
\label{exp}
\feqr

\noi The coefficients of the Taylor expansion are tensors and can be expressed in terms of the components $R^{l}_{m n p}$ 
\footnote{Our conventions are: $R^{l}_{mnp}= \partial_{n} \Gamma^{l}_{mp} + \Gamma^{k}_{mp}\Gamma^{l}_{kn} - 
(n \leftrightarrow p)$, $R_{\mu \nu}=R^{\alpha}_{\,\,\, \mu \alpha \nu}$ and $R=R^{\mu}_{\,\,\mu}$.} of
 the Riemann curvature tensor and the covariant derivatives $D_{k} T_{lm}$ and $D_{k}R^{l}_{m n p}$. One without 
much effort can prove that:
\beqr
\partial_{(i_1} \partial_{i_2} \cdots \partial_{i_{n-1}} 
\bar{\Gamma}^{l}_{i_n ) k} = &-& \left(\frac{n-1}{n+1} \right) \Big[
D_{(i_1} D_{i_2} \cdots D_{i_{n-2}} \bar{R}^{l}_{i_{n-1}k i_n)} \non \\
&+& \partial_{(i_1} \partial_{i_2} \cdots \partial_{i_{n-2}} \left
(\bar{\Gamma}^{\alpha}_{i_{n-1}k} \bar{\Gamma}^{l}_{\alpha i_n )} 
\right) \non \\
&-& \partial_{(i_1} \partial_{i_2} \cdots \partial_{i_{n-3}}
\left(\bar{\Gamma}^{l}_{i_{n-2} \alpha} \bar{R}^{\alpha}_{i_{n-1}ki_n )} 
- l \leftrightarrow \alpha, \alpha \leftrightarrow k \right) \non \\
&-& \partial_{(i_1} \partial_{i_2} \cdots \partial_{i_{n-4}}
\left(\bar{\Gamma}^{l}_{i_{n-3} \alpha} D_{i_{n-2}}
\bar{R}^{\alpha}_{i_{n-1}ki_n )} 
- l \leftrightarrow \alpha, \alpha \leftrightarrow k \right) \non \\
&\vdots& \non \\
&-& \left( \partial_{(i_1} \bar{\Gamma}^{l}_{i_2 \alpha} D_{i_3} \cdots D_{i_{n-2}}
\bar{R}^{\alpha}_{i_{n-1} k i_n )} - l \leftrightarrow \alpha, \alpha \leftrightarrow k
\right) \Big]
\label{genex}
\feqr

\noi where the interchange of covariant and contravariant indices act independently. Expression\rf{genex} reproduces 
for various values of n\footnote{In \cite{alvarez} there is a misprint for the $n=4$ case. A minus sign is needed in 
front of the $\frac{2}{9}$-term.} the following results:
\beqr
\partial_{(i_1}\bar{\Gamma}^{l}_{i_2 ) k} = & & \frac{1}{3}
\bar{R}^{l}_{(i_1 i_2) k} \non \\
\partial_{(i_1}\partial_{i_2}\bar{\Gamma}^{l}_{i_3) k} = 
&-&\frac{1}{2}D_{(i_1}\bar{R}^{l}_{i_2 k i_3 )} \non \\
\partial_{(i_1}\partial_{i_2} \partial_{i_3}
\bar{\Gamma}^{l}_{i_4) k} = 
&-&\frac{3}{5} \left[D_{(i_1} D_{i_2} \bar{R}^{l}_{i_3 k i_4 )}
+ \frac{2}{9} \bar{R}^{l}_{(i_1 i_2 \alpha} \bar{R}^{\alpha}_{i_3 i_4 )k}
\right] \non \\
\partial_{( i_1} \partial_{i_2} \partial_{i_3} \partial_{i_4}
\bar{\Gamma}^{l}_{i_5) k} =
&-&\frac{2}{3} \left[D_{(i_1} D_{i_2} D_{i_3} \bar{R}^{l}_{i_4 k i_5 )}
- D_{(i_1}\bar{R}^{\alpha}_{i_2 k i_3} \bar{R}^{l}_{i_4 i_5 )\alpha} 
\right]\non \\
\partial_{( i_1} \partial_{i_2} \partial_{i_3} \partial_{i_4}
\partial_{i_5} \bar{\Gamma}^{l}_{i_6) k} =  
&-&\frac{5}{7} \Big[D_{(i_1} \cdots D_{i_4} \bar{R}^{l}_{i_5 k i_6 )} \non \\
&-& \frac{1}{5} \left(7D_{(i_1}D_{i_2}\bar{R}^{l}_{i_3 \alpha i_4} 
\bar{R}^{\alpha}_{i_5 i_6 )k } + D_{(i_1}D_{i_2}\bar{R}^{\alpha}_{i_3 k i_4} 
\bar{R}^{l}_{i_5 i_6 ) \alpha } \right) \non \\
&+& \frac{3}{2}D_{(i_1} \bar{R}^{\alpha}_{i_2 k i_3}
 D_{i_4} \bar{R}^{l}_{i_5 \alpha i_6 )} - \frac{16}{45}
\bar{R}^{l}_{(i_1 i_2 \alpha} \bar{R}^{\alpha}_{i_3 i_4 \beta}
\bar{R}^{\beta}_{i_5 i_6 )k}\Big] \non \\
\cdots
\label{resgen}
\feqr 

\noi The coefficients of\rf{exp} can be rewritten as:
\beqr
\partial_{( i_1} \partial_{i_2} \cdots \partial_{i_n )} \bar{T}_{k_1 k_2} =
& & D_{(i_1}D_{i_2} \cdots D_{i_n )}\bar{T}_{k_1 k_2} \non \\ 
&+& \partial_{( i_1} \partial_{i_2} \cdots \partial_{i_{n-1} )}
\left[\bar{\Gamma}^{\alpha}_{i_n k_1 )} \bar{T}_{\alpha k_2} + 
k_1 \leftrightarrow k_2 \right] \non \\
&+& \partial_{( i_1} \partial_{i_2} \cdots \partial_{i_{n-2} )}
\left[\bar{\Gamma}^{\alpha}_{i_{n-1} k_1} D_{i_n )}\bar{T}_{\alpha k_2} + 
k_1 \leftrightarrow k_2 \right] \non \\
&\vdots& \non \\
&+& \partial_{( i_1} \partial_{i_2}
\left[\bar{\Gamma}^{\alpha}_{i_3 k_1} D_{i_4} \cdots D_{i_n )}
\bar{T}_{\alpha k_2} + 
k_1 \leftrightarrow k_2 \right] \non \\
&+& \partial_{( i_1}
\left[\bar{\Gamma}^{\alpha}_{i_2 k_1} D_{i_3} \cdots D_{i_n )}
\bar{T}_{\alpha k_2} + k_1 \leftrightarrow k_2 \right]. 
\label{coeexp}
\feqr

\noi Expressions\rf{genex} and\rf{coeexp} compose the building blocks of the current recursive method which produces 
the following results for different values of n:
\beqr
\partial_{( i_1} \partial_{i_2 )} \bar{T}_{k_1 k_2} &=& 
D_{(i_1} D_{i_2)} \bar{T}_{k_1 k_2} - \frac{1}{3} \left[
\bar{R}^{\rho}_{(i_1 k_1 i_2 )} \bar{T}_{\rho k_2} + 
k_1 \leftrightarrow k_2 \right] \non \\
\partial_{( i_1} \partial_{i_2} \partial_{i_3 )} \bar{T}_{k_1 k_2} = & & 
D_{(i_1} D_{i_2} D_{i_3 )}\bar{T}_{k_1 k_2} \non \\
&+& \left[\partial_{(i_1} \partial_{i_2} \bar{\Gamma}^{\rho}_{i_3 ) k_1} 
\bar{T}_{\rho k_2} + 2 \partial_{(i_1} \bar{\Gamma}^{\rho}_{i_2 k_1}
D_{i_3 )} \bar{T}_{\rho k_2} + k_1 \leftrightarrow k_2\right] \non \\
&-& \frac{1}{3} \left(\bar{R}^{\rho}_{(i_1 k_1 i_2} D_{i_3 )} 
\bar{T}_{\rho k_2} + k_1 \leftrightarrow k_2 \right) \non \\
\partial_{( i_1} \partial_{i_2} \partial_{i_3} \partial{i_4 )} \bar{T}_{k_1 k_2} = & & 
D_{(i_1} D_{i_2} D_{i_3} D_{i_4 )} \bar{T}_{k_1 k_2} \non \\
&+& \Big[\partial_{( i_1} \partial_{i_2} \partial_{i_3} \bar{\Gamma}^{\rho}_{i_4 ) k_1} \bar{T}_{\rho k_2}
+ 3 \partial_{( i_1} \partial_{i_2} \bar{\Gamma}^{\rho}_{i_3  k_1}D_{i_4 )} \bar{T}_{\rho k_2} \non \\
&+& 3\partial_{(i_1} \bar{\Gamma}^{\rho}_{i_2 k_1} \left(D_{i_3}D_{i_4 )} \bar{T}_{\rho k_2} - \frac{1}{3} 
\left(\bar{R}^{\sigma}_{i_3 \rho i_4 )} \bar{T}_{\sigma k_2} + \rho \leftrightarrow k_2 \right)\right) +
k_1 \leftrightarrow k_2 \Big] \non \\
\partial_{( i_1} \partial_{i_2} \partial_{i_3} \partial{i_4} \partial_{i_5 )} \bar{T}_{k_1 k_2} = & & 
D_{(i_1}D_{i_2}D_{i_3}D_{i_4}D_{i_5 )} \bar{T}_{k_1 k_2} \non \\
&-& \Big[ \frac{10}{3} \bar{R}^{\alpha}_{(i_1 k_1 i_2} D_{i_3} \cdots D_{i_5 )} \bar{T}_{\alpha k_2} +
5 D_{(i_1}\bar{R}^{\alpha}_{i_2 k_1 i_3}D_{i_4}D_{i_5)}\bar{T}_{\alpha k_2} \non \\
&+& 3D_{(i_1}D_{i_2}\bar{R}^{\alpha}_{i_3 k_1 i_4} D_{i_5)}\bar{T}_{\alpha k_2} + 
\frac{2}{3} D_{(i_1}D_{i_2}D_{i_3} \bar{R}^{\alpha}_{i_4 k_1 i_5)} \bar{T}_{\alpha k_2} \non \\
&-& \frac{2}{3}D_{(i_1} \bar{R}^{\rho}_{i_2 k_1 i_3} \bar{R}^{\alpha}_{i_4 i_5 )\rho} \bar{T}_{\alpha k_2}
+D_{(i_1}\bar{R}^{\alpha}_{i_2 k_1 i_3} \left(\bar{R}^{\rho}_{i_4 \alpha i_5)} \bar{T}_{\rho k_2} + 
\alpha \leftrightarrow k_2 \right) \non \\
&+& \frac{2}{3} \left(D_{(i_1}\bar{R}^{\rho}_{i_2 \alpha i_3} \bar{T}_{\rho k_2}+ \alpha \leftrightarrow k_2 \right)
\bar{R}^{\alpha}_{i_4 i_5) k_1} \pm k_1 \leftrightarrow k_2\Big] \non \\
\partial_{( i_1} \partial_{i_2} \partial_{i_3} \partial{i_4} \partial_{i_5} \partial_{i_6)}\bar{T}_{k_1 k_2} = & & 
D_{(i_1} \cdots D_{i_6)} \bar{T}_{k_1 k_2} \non \\
&+& \Big[ \partial_{( i_1} \cdots \partial_{i_5} \bar{\Gamma}^{\alpha}_{i_6) k_1} \bar{T}_{\alpha k_2} 
+ 6  \partial_{( i_1} \cdots \partial_{i_4} \bar{\Gamma}^{\alpha}_{i_5 k_1} D_{i_6)}\bar{T}_{\alpha k_2} \non \\
&+& 15\partial_{( i_1} \cdots \partial_{i_3} \bar{\Gamma}^{\alpha}_{i_4 k_1} D_{i_5}D_{i_6)}\bar{T}_{\alpha k_2} 
+20 \partial_{( i_1} \partial_{i_2} \bar{\Gamma}^{\alpha}_{i_3 k_1} D_{i_4} \cdots D_{i_6)}\bar{T}_{\alpha k_2} \non \\
&+& 14\partial_{( i_1} \bar{\Gamma}^{\alpha}_{i_2 k_1} D_{i_3} \cdots D_{i_6)}\bar{T}_{\alpha k_2} \non \\
&+& 10 \partial_{( i_1} \cdots \partial_{i_3} \bar{\Gamma}^{\alpha}_{i_4 k_1} \left(\partial_{i_5} 
\bar{\Gamma}^{\rho}_{i_6) \alpha} \bar{T}_{\rho k_2} + \alpha \leftrightarrow k_2 \right) \non \\
&+& 10 \partial_{( i_1} \partial_{i_2} \bar{\Gamma}^{\alpha}_{i_3 k_1} \left(\partial_{i_4} \partial_{i_5} 
\bar{\Gamma}^{\rho}_{i_6) \alpha} \bar{T}_{\rho k_2} + \alpha \leftrightarrow k_2 \right) \non \\
&+& 36\partial_{( i_1} \partial_{i_2} \bar{\Gamma}^{\alpha}_{i_3 k_1} \left(\partial_{i_4} 
\bar{\Gamma}^{\rho}_{i_5 \alpha} D_{i_6)}\bar{T}_{\rho k_2} + \alpha \leftrightarrow k_2 \right) \non \\
&+& 5 \partial_{( i_1} \bar{\Gamma}^{\alpha}_{i_2 k_1} \left(\partial_{i_3} \cdots \partial_{i_5}
\bar{\Gamma}^{\rho}_{i_6) \alpha} \bar{T}_{\rho k_2} + \alpha \leftrightarrow k_2 \right) \non \\
&+& 24 \partial_{( i_1} \bar{\Gamma}^{\alpha}_{i_2 k_1} \left(\partial_{i_3} \partial_{i_4} 
\bar{\Gamma}^{\rho}_{i_5 \alpha}D_{i_6)}\bar{T}_{\rho k_2} + \alpha \leftrightarrow k_2 \right) \non \\
&+& 45 \partial_{( i_1} \bar{\Gamma}^{\alpha}_{i_2 k_1} \left(\partial_{i_3} \bar{\Gamma}^{\rho}_{i_4 \alpha}
D_{i_5} D_{i_6)}\bar{T}_{\rho k_2} + \alpha \leftrightarrow k_2 \right) \non \\ 
&+& 15 \left[ \partial_{(i_1} \bar{\Gamma}^{\alpha}_{i_2 k_1} \partial_{i_3} \bar{\Gamma}^{\rho}_{i_4 \alpha}\left( 
\partial_{i_5} \bar{\Gamma}^{\sigma}_{i_6) \rho} \bar{T}_{\sigma k_2} + \rho \leftrightarrow k_2 \right) 
+ \alpha \leftrightarrow k_2 \right] \non \\
&+& \partial_{(i_1} \bar{\Gamma}^{\alpha}_{i_2 k_1} D_{i_3} \cdots D_{i_6)} \bar{T}_{\alpha k_2}
+ k_1 \leftrightarrow k_2 \Big].
\label{coedn}
\feqr

\noi If the second rank tensor with components $\bar{T}_{k_1 k_2}$ is replaced by the metric components 
$\bar{g}_{k_1 k_2}$ then the related covariant derivatives (provided we deal with a torsion free affine connection) 
vanish and the above expressions are simplified. One could derive for $n=5$ the result:
\beqr
\partial_{(i_1} \cdots \partial_{i_5)}\bar{g}_{k_1 k_2} = \frac{4}{3} \left[D_{i_1} \cdots D_{i_3} 
\bar{R}_{k_1 i_4 i_5 k_2} 
+ 2 \left(D_{i_1} \bar{R}_{k_1 i_2 i_3 \rho} \bar{R}^{\rho}_{i_4 i_5 k_2} +  k_1 \leftrightarrow k_2 \right) 
\right].
\label{odty}
\feqr

\noi On the other hand for $n=6$ one gets:
\beqr
\partial_{( i_1} \partial_{i_2} \partial_{i_3} \partial{i_4} \partial_{i_5} \partial_{i_6)}\bar{g}_{k_1 k_2} = & &
\frac{10}{7}D_{(i_1} \cdots D_{i_4} \bar{R}_{k_1 i_5 i_6) k_2} + \frac{34}{7}
\left(D_{(i_1}D_{i_2} \bar{R}_{k_1 i_3 i_4 \rho} 
\bar{R}^{\rho}_{i_5 i_6) k_2} + k_1 \leftrightarrow k_2 \right) \non \\
&+& \frac{55}{7} D_{(i_1} \bar{R}_{k_1 i_2 i_3 \rho} D_{i_4} \bar{R}^{\rho}_{i_5 i_6) k_2} +
\frac{16}{7} \bar{R}_{k_1 (i_1 i_2 \rho} \bar{R}^{\rho}_{i_3 i_4 l} \bar{R}^{l}_{i_5 i_6) k_2}.
\label{evety}
\feqr

\noi Thus, plugging into\rf{exp} expressions\rf{odty} and\rf{evety} we end up with the following expansion of the 
metric tensor in RNC:
\beqr
g_{k_1 k_2} = & & \bar{g}_{k_1 k_2} + \frac{1}{2!} \frac{2}{3}
\bar{R}_{k_1 i_1 i_2 k_2} \xi^{i_1} \xi^{i_2} \non \\
&+& \frac{1}{3!} D_{i_1} \bar{R}_{k_1 i_2 i_3 k_2} \xi^{i_1} \cdots \xi^{i_3} \non \\
&+& \frac{1}{4!}\frac{6}{5} \left[D_{i_1}D_{i_2} \bar{R}_{k_1 i_3 i_4 k_2} 
+ \frac{8}{9} \bar{R}_{k_1 i_1 i_2 m} \bar{R}^{m}_{i_3 i_4 k_2}\right] \xi^{i_1} \cdots \xi^{i_4}\non \\
&+& \frac{1}{5!} \frac{4}{3} \left[D_{i_1} \cdots D_{i_3} \bar{R}_{k_1 i_4 i_5 k_2} 
+ 2 \left(D_{i_1} \bar{R}_{k_1 i_2 i_3 \rho} \bar{R}^{\rho}_{i_4 i_5 k_2} +  k_1 \leftrightarrow k_2 \right) 
\right] \xi^{i_1} \cdots \xi^{i_5} \non \\
&+& \frac{1}{6!} \frac{10}{7}\Big[ D_{i_1} \cdots D_{i_4} \bar{R}_{k_1 i_5 i_6 k_2} + \frac{17}{5}\left(D_{i_1}D_{i_2} 
\bar{R}_{k_1 i_3 i_4 \rho} 
\bar{R}^{\rho}_{i_5 i_6 k_2} + k_1 \leftrightarrow k_2 \right) \non \\
&+& \frac{11}{2}D_{i_1} \bar{R}_{k_1 i_2 i_3 \rho} D_{i_4} \bar{R}^{\rho}_{i_5 i_6 k_2} +
\frac{8}{5}\bar{R}_{k_1 i_1 i_2 \rho} \bar{R}^{\rho}_{i_3 i_4 l} \bar{R}^{l}_{i_5 i_6 k_2} \Big] 
\xi^{i_1} \cdots \xi^{i_6}\non \\
&+& O(\xi^{i_1} \cdots \xi^{i_7}).
\label{exme}
\feqr

\noi The expansion\rf{exme} is in perfect agreement with that quoted in \cite{schubert}. The inverse of the metric tensor
(obeying $g_{k_1 k_2} g^{k_2 k_3} = \delta^{k_1}_{k_3}$) can be found to be:
\beqr
g^{k_1 k_2} = & & \bar{g}^{k_1 k_2} - \frac{1}{2!} \frac{2}{3}
\bar{R}^{\, k_1 \quad k_2}_{\,\,\,\,\,\,  i_1 i_2} \xi^{i_1} \xi^{i_2} \non \\
&-& \frac{1}{3!} D_{i_1} \bar{R}^{\, k_1 \quad k_2}_{\,\,\,\,\,\,  i_2 i_3} \xi^{i_1} \cdots \xi^{i_3} \non \\
&-& \frac{1}{4!}\frac{6}{5} \left[D_{i_1}D_{i_2} \bar{R}^{\, k_1 \quad k_2}_{\,\,\,\,\,\,  i_3 i_4} 
- \frac{4}{3} \bar{R}^{k_1}_{\, i_1 i_2 m} \bar{R}^{\, m \quad k_2}_{\,\,\,\,\,\, i_3 i_4}\right] 
\xi^{i_1} \cdots \xi^{i_4}\non \\
&-& \frac{1}{5!} \frac{4}{3} \left[D_{i_1} \cdots D_{i_3} \bar{R}^{\, k_1 \quad k_2}_{\,\,\,\,\,\, i_4 i_5} 
- 3 \left(D_{i_1} \bar{R}^{k_1}_{\, i_2 i_3 \rho} \bar{R}^{\, \rho \quad k_2}_{\,\,\,\, i_4 i_5} +  
k_1 \leftrightarrow k_2 \right) \right] \xi^{i_1} \cdots \xi^{i_5} \non \\
&-& \frac{1}{6!} \frac{10}{7}\Big[ D_{i_1} \cdots D_{i_4} \bar{R}^{\, k_1 \quad k_2}_{\,\,\,\,\,\, i_5 i_6} 
- 5\left(D_{i_1}D_{i_2} \bar{R}^{k_1}_{\, i_3 i_4 \rho} 
\bar{R}^{\, \rho \quad k_2}_{\,\,\,\, i_5 i_6} + k_1 \leftrightarrow k_2 \right) \non \\
&-& \frac{17}{2}D_{i_1} \bar{R}^{k_1}_{i_2 i_3 \rho} D_{i_4} \bar{R}^{\rho \quad k_2}_{\,\,\,\, i_5 i_6} +
\frac{16}{3}\bar{R}^{k_1}_{i_1 i_2 \rho} \bar{R}^{\rho }_{i_3 i_4 l} 
\bar{R}^{l \quad k_2}_{i_5 i_6} \Big] \xi^{i_1} \cdots \xi^{i_6}\non \\
&+& O(\xi^{i_1} \cdots \xi^{i_7}).
\label{inmet}
\feqr

As a simple check one could consider the symmetric space $V_n$ in RNC for which $D_{k}\bar{R}_{lmnp} =0$ and prove that indeed
 the R terms in\rf{exme} satisfy:
\beqr
g_{\alpha \beta} = \bar{g}_{\alpha \beta} + \frac{1}{2} \sum_{k=1}^{\infty} (-1)^h \frac{2^{2k+2}}{(2k+2)!}
f^{\sigma_1}_{\alpha}f^{\sigma_1}_{\sigma_2} \cdots f_{\sigma_{k-1} \beta} ,\quad h= \left\{ \begin{array}{ll}
k+1 & \textit{if k is even} \\
k   & \textit{if k is odd}
\end{array}
\right.
\label{symmsp}  
\feqr

\noi where $f^{\sigma_1}_{\alpha}= \bar{R}_{\alpha i_1 i_2}^{\sigma_1} \xi^{i_1} \xi^{i_2}$ 
and $f_{\sigma_{k-1} \beta} = \bar{R}_{i_p \sigma_{k-1} i_{p+1} \beta} \xi^{i_p} \xi^{i_{p+1}}$.

\section{Conclusions}

We have shown the method one could rely on to evaluate the expansion of the components of the metric tensor in RNC at 
a specific order. The recursive structure permits an answer the derivation of which becomes cumbersome when one 
attempts to calculate higher order terms.    

\bibliographystyle{plain}

\end{document}